\documentclass[12pt,a4paper]{article}
\usepackage[utf8]{inputenc}
\usepackage[left=3cm,right=3cm,top=3cm,bottom=3cm]{geometry}
\usepackage{graphicx} % Required for inserting images
\usepackage{xurl}
\usepackage{comment}
\usepackage{setspace}
\usepackage{ragged2e}
\usepackage{booktabs}
\usepackage{bbm}
\usepackage{amsmath}
\usepackage{amssymb}
\usepackage{placeins}
\usepackage[capposition=bottom]{floatrow}
\usepackage[dvipsnames]{xcolor}
\usepackage[colorlinks=true]{hyperref}
	\hypersetup{citecolor=MidnightBlue, anchorcolor=Black, urlcolor=MidnightBlue, linkcolor=MidnightBlue}
\usepackage[nottoc]{tocbibind}
\usepackage{tikz}
\usepackage{pdflscape} % or {lscape}
\usepackage{longtable}
\usepackage{verbatim}

\usepackage[backend=biber, style=apa, bibstyle=authoryear, dashed = false, maxcitenames = 2, maxbibnames=9, giveninits=true, sorting=nyt, date=year, eprint = false, doi=false, isbn=false]{biblatex} 
\newcommand\fnote[1]{%
    \vspace{-0.5\baselineskip} % Adjust vertical space if needed
    \parbox{\linewidth}{\footnotesize#1}\vspace{0.5\baselineskip} % Put the note in a box with smaller font
}

\title{
Academic Knowledge: Does it Reflect the Combinatorial Growth of Technology?%\footnote{I am grateful for valuable feedback from Ulrich Herb and Maikel Pellens. All errors and opinions are solely my own.}
} % I thank XXX for helpful feedback.
\author{
W.~Benedikt Schmal\thanks{Ilmenau University of Technology, Economic Theory Group. Ehrenbergstr.~29, 98693 Ilmenau, Germany \& KU Leuven University, Department for Management, Strategy, and Innovation (MSI). Namsestraat 69, 3000 Leuven, Belgium. Email: \href{mailto:wolfgang-benedikt.schmal@tu-ilmenau.de}{wolfgang-benedikt.schmal@tu-ilmenau.de}\\ \textbf{ORCiD}: \href{https://orcid.org/0000-0003-2400-2468}{0000-0003-2400-2468}.}\\
{\normalsize TU Ilmenau \& MSI, KU Leuven}\\
}
\date{\today}

\begin{document}

\maketitle

\begin{onehalfspace}
    \begin{abstract}
    \noindent I explore the concept of growth being rooted in the recombination of existing technology as an explanation for the remarkable growth witnessed during the Industrial Revolution as it was recently proposed by \textcite{Koppl.2023}. I adapt their combinatorial growth theory to assess its applicability in generating academic knowledge within universities and research institutions, particularly in the field of economics. The central question is whether significant combinatorial growth can also be anticipated in academia. The current career structures discourage the recombination of ideas, theories, or methods, making it more advantageous for early career researchers to stick to the status quo. I employ machine-learning-based natural language analysis of the top 5 journals in economics. The analysis reveals limited correlations between topics over the past three decades, suggesting the presence of isolated topic islands rather than productive recombination. This confirms the theoretical considerations beforehand. Overall, the institutional order of academia makes combinatorial growth at the research frontier unlikely.
\end{abstract}
\end{onehalfspace}

\newpage
\begin{doublespace}
    
\section{Introduction}
\label{sec.intro}

The Industrial Revolution and how it triggered massive economic growth over the following decades, if not centuries, continues to be an economic miracle. It is widely acknowledged that many factors came together and contributed to the greatest creation of wealth and well-being in human history. %Due to the coincidence of a plethora of drivers, it is difficult for economists to net out specific reasons. 
In a new contribution in the field of complexity economics, \textcite{Koppl.2023} argue that it was also or mainly due to the fruitful recombination of existing technologies that economic growth fluctuated around zero during medieval times and accelerated so much during the industrial revolution until today. Doing so, they follow a similar approach by \textcite{Arthur.2009}, who emphasizes that every technology consists of a combination of several other technologies, e.g., a car that combines thousands of individual technological features that form -- in combination -- a new technological good, namely the car.

This paper tries to apply the combinatorial approach of \textcite{Koppl.2023} as an explanation for economic growth, especially the sudden and rapid growth since the Industrial Revolution, to the sphere of generating academic knowledge. The paper particularly discusses the academic discipline of economics. Nevertheless, many insights are also applicable to other fields of research. The generation of \emph{novel} research is of utmost importance for progress in multitudinous dimensions. However, the example of economics especially demonstrates that career incentives are biased in a way that novel, interdisciplinary research is often inferior in generating attractive career prospects than technically sophisticated but incremental contributions to the literature. 

I point to three reasons: increasing search costs, high opportunity costs of intellectual diversity, and choice overload. Lastly, I also discuss the negative effects of increasing diversification and the formation of academic niches within a discipline \parencite[see, e.g.,][]{Visser.2023}. All four aspects disincentivize the rational (economics) researcher from exploring recombinations but guide them towards providing more of the same at the cost of combinatorial growth. In light of a missing counterfactual, it is barely possible to pin down the opportunity costs of this suboptimal path. I amend my analysis by setting up a structural topic model. This machine-learning technique from the realm of natural language processing (NLP) enables me to detect the underlying topics of the papers published in the so-called `top 5' journals in economics. Doing so, I detect a notable lack of correlations between these topics. This hints at a high prevalence of isolated topic `islands' in the leading journals of the discipline, which does not tie in with the idea of combinatorial growth. 

%The cornerstone of this paper is the book \emph{`Explaining Technology'} by \textcite{Koppl.2023}. 
In light of the fact that the present paper relates the concept presented by \textcite{Koppl.2023} to academic research, there exists a body of research that examines the growth in academia analogous to overall economic growth.\footnote{In terms of overall economic growth, \textcite{Koppl.2023} aim at deepening our understanding in addition to seminal work on the role of institutions by \textcite{Acemoglu.2001, Acemoglu.2005},  \textcite{Eichengreen.1994} or evolution by \textcite{Galor.2001}. However, this is not the focus of the present paper.} \textcite{Bornmann.2015} observe since the 1980s an upward trend in academic publications that seems to follow an exponential pattern. More importantly, they are able to look at citations or else references from one article to another. There, they can go back to the early 1600s. Their empirical analysis demonstrates a remarkably exponential growth in citations and references in academic work since then.

NLP techniques gained traction in economics recently. As it turns textual into quantitative data, it allows the exploitation of a plethora of resources that are based in natural language \parencite{Gentzkow.2019}. It has been particularly strong in analyzing central bank communication \parencite[see, e.g,][]{Hansen.2018, Ferrara.2022, Kusters.2022}. However, it also has already been applied to study research contents \parencite[see, e.g.,][]{Ebadi.2021, Schmal.2023} or corporate communication \parencite{Kumar.2022}.

The remainder of the paper is structured as follows. In Section \ref{sec.combi}, I discuss how far the combinatorial concept of growth proposed by \textcite{Arthur.2009} and \textcite{Koppl.2023} can be related to the academic sphere. Particularly, I examine the issues of search costs, intellectual diversity, and choice overload. In Section \ref{sec.top5}, I apply a machine-learning-based topic model to examine the themes of the leading `top 5' journals in economics and to which extent recombinations are likely. Section \ref{sec.conc} discusses the main findings and concludes.

\section{A Combinatorial Understanding of Research}
\label{sec.combi}

The core of the approach by \textcite{Koppl.2023} forms the understanding of economic growth as a combinatorial process. In particular, the authors propose that every good is either a combination of other good or a modification of a previous good. They use an axe as an example, where newer axes may be of better quality or sharpness, but there is no recombination of goods in producing them (see p.~12, ibid). They further note that not every recombination is, at least initially, useful or well-designed. Rather, ugly combinations may occur, which either get improved or disappear again. The combinatorial power arises from the increase in goods, which allows for an exponential growth in potential combinations. 

Interestingly, we also observe exponential growth in academic research. As mentioned beforehand, \textcite{Bornmann.2015} document exponential growth in the number of references in academic publications. It remarkably mirrors the concept of recombination as citations, much more than publications, capture the connection of new to earlier work. It does not necessarily imply recombination in a technological sense but could simply be a referral to similar or, on the other hand, contradicting work, which is rarely found in technology. Nevertheless, it ties in with the concept of \textcite{Koppl.2023}, simply because more work researchers can build upon implies a broader set of philosophical, conceptual, and methodological approaches they can choose from and which they can recombine for their own research purposes. 

While the observation of massively growing references over the past centuries implies a surprisingly parallel development between technology and academic science, there exists a stark contrast in career mechanisms, which suggests a weaker continued growth than in the technosphere. \textcite{Heckman.2020} document what they call the `tyranny of the top 5' in economics, i.e., a strong dominance of the five journals \emph{American Economic Review, Econometrica, Journal of Political Economy, Quarterly Journal of Economics,} and \emph{Review of Economic Studies}, which are considered the stellar and most important journals that also determine career paths. More generally, journal reputation in economics is highly convex, which means that one publication in an outlet perceived as a `top journal' outweighs plenty of publications in what is perceived as weaker journals by the discipline \parencite{Brauninger.2003, Haucap.2015}. 

\textcite{Schmal.2023} already documents for the topic of collusion in the subfield of industrial economics a worrying streamlining of research on cartels. Notably, he detects that researchers endogenously narrow the set of topics covered in more recent papers based on varying citation rates across topics in older publications. It ties in with the problem that has been coined `publish or perish,' i.e., the necessity to publish one's own work as often as possible in the best possible journals. This leads to high pressure on academics, incentivizes academic fraud, and cuts down diversity in research interests \parencite{vanDalen.2012, vanDalen.2021}. For example, policy contributions are, especially for economists, who would be able to comment a lot on current political affairs, virtually worthless.

\begin{figure}[htbp]
    \centering
\begin{tikzpicture}[scale=1.2] 
    % Coordinates for the left triangle (vertex at the top)
    \coordinate (A) at (0,0);  % Bottom-left corner
    \coordinate (B) at (4,0);  % Bottom-right corner
    \coordinate (C) at (2,4);  % Top vertex
    % Draw the first triangle
    \draw[thick] (A) -- (B) -- (C) -- cycle;
    % Coordinates for the second triangle
    \coordinate (D) at (8,0);  % Bottom vertex
    \coordinate (E) at (6,4);  % Top-left corner
    \coordinate (F) at (10,4);  % Top-right corner
    % Draw the second triangle
    \draw[thick] (D) -- (E) -- (F) -- cycle;
    % Labels for the left triangle
    \node at (2,4.5) {\textbf{Editors}};    % Label above the left triangle
    \node at (2,-0.5) {\textbf{Scientists}};  % Label below the left triangle
    % Labels for the right triangle
    \node at (8,4.5) {\textbf{Consumers}};  % Label above the right triangle
    \node at (8,-0.5) {\textbf{Entrepreneurs}};% Label below the right triangle
\end{tikzpicture}
    \caption{Differences between researchsphere and technosphere}
    \label{fig:1}
\end{figure}
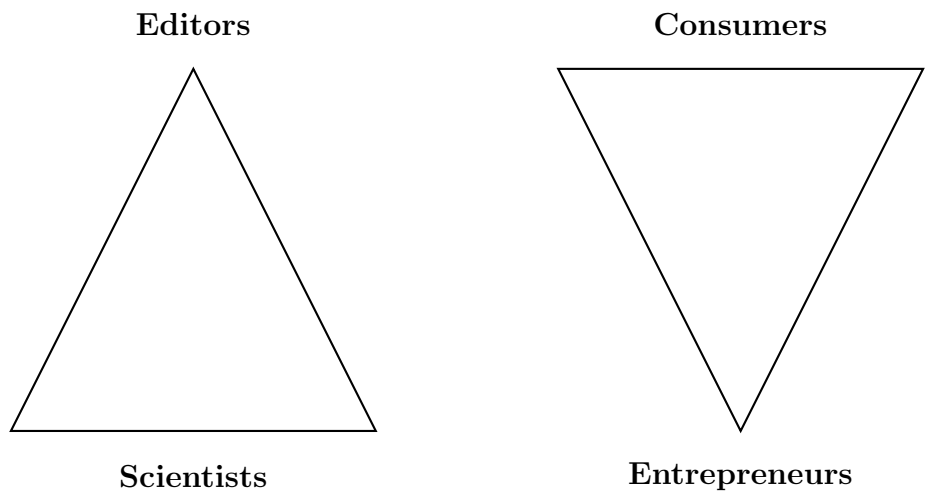

The second issue that makes continued combinatorial growth in academia unlikely is the difference between `researchsphere' and technosphere depicted in Figure \ref{fig:1}. While a group of entrepreneurs developing a new good or service finds themselves able to serve a plethora of consumers with a broad set of preferences, tastes, and utility functions, scientists are in the opposite setting. Here, a plethora of scientists try to attract a small group of journal editors, namely those serving on the editorial boards of the top journals. 
Put differently, many researchers compete for the scarce space in a few journals. Thus, it is rational for them to figure out the research interests of these editors and adjust their own research to these topics.\footnote{Even more beneficial are personal ties to some editors as shown by \textcite{Colussi.2018, Ductor.2022}.} \textcite{Visser.2023} as well as \textcite{Svorencik.2023} document for the leading journals and the American Economic Association -- globally the most influential academic association of the discipline -- long-standing personal and geographical continuities. This, however, weakens ceteris paribus scientific progress as researchers aligning their work with those of the editors may no longer work on their best ideas and combine existing work they consider the most promising from a general perspective but try to mimic current publications in those journals. This concern is not new as already \textcite{Hodgson.1999} were worried that the high degree of editor concentration at top journals and (US) top departments could be detrimental to innovation. \textcite{Heckman.2020} echo this concern approximately two decades later while putting a specific emphasis on the editorial boards of the top 5 journals in economics. 

Of course, there exists a mass of `me-too' inventions in the technosphere that attempt to copy existing `combinations.' However, pinchbecks can only be successful if they increase the welfare of a sufficient number of consumers. In the case of imitating products, this would be usually the price. But again, entrepreneurs in the technosphere need to convince a critical mass of individuals, while in academia, it is to the benevolence of a single editor and (at most) a handful of reviewers. \\ 

\noindent \textbf{Search Costs:} Both the techno- and the researchsphere face search costs. The more goods and services that are offered, the higher the effort to review the whole menu. This holds also for online shopping. For prices, comparison engines exist a lot, but if one wants to compare product features aside from the price, consumers need to review many goods manually. Search costs are also a reason why the law of one price often does not hold and why price dispersion even for homogeneous goods exists \parencite[see, e.g.,][]{Salop.1977, Varian.1980}. Interestingly, \textcite{Stahl.1989} proves that for sequential search -- a consumer searches one (online) shop after another -- an \emph{increase} in shops, which would, in plain theory, lead to more competition and lower prices, can lead to higher prices. 

How can we relate this to the research sphere? With the permanent growth in academic publications, researchers face the challenge of an evergrowing body of literature in their own field as well as all other fields. To discover novel techniques heretofore unused in a specific discipline, a curious researcher faces growing fixed set-up costs to get familiar with another strand of the literature. This does not necessarily point at a fully different discipline, say economics and paleontology; it can already be remote subfields within one discipline. Put differently, the growth in the research body also implies a growth in search costs in the researchsphere. Other than in the technosphere, a researcher cannot simply quit searching once they find something interesting to them. They have to review also the internal reception and usage of a technique they uncovered before applying it to their main field of research. \\

\noindent \textbf{Intellectual Diversity:} But even if a researcher incurs the costs of exploring another field, they still face the challenge of peer review: Once one submits a draft to an academic journal to get it published there, the handling editor -- conditional on the draft surviving the initial desk rejection hurdle -- usually approaches one to four anonymous referees that carefully read and comment on the draft. Here, interdisciplinarity can become quite problematic. If we assume that the author successfully recombined two or more existing approaches in order to create something new in the spirit of \textcite{Koppl.2023}, they are necessarily unique in the sense that no one else has worked on exactly that combination. As a consequence, editors will face challenges in finding reviewers with fitting interests. 

It is likely that they will ask reviewers who are familiar with one or less than all fields of research addressed in the submitted paper. Thus, every referee will face the problem that they are unfamiliar with one or more topics of such a recombined interdisciplinary paper. As a consequence, they will either criticize the combination of methods/topics/ideas and give bad feedback or no feedback on the combination. Thus, the paper will have problems getting accepted.  but even if it gets accepted, the quality should be, ceteris paribus, worse in light of the fact that referee reports may not be as helpful as in the case of fully informed reviewers. This leads to a comparative disadvantage of papers that apply novel recombinations.

Lastly, researchers could go one step further: Instead of obtaining knowledge of other fields all by themselves, they could join forces with fellow scientists from other disciplines, e.g., at their university. This lowers the set-up costs as they could get experts on the planned recombinations on board. But by doing so, they would face another, already sketched issue of the researchsphere that makes a combinatoric revolution in the future unlikely: The career mechanisms. Different disciplines have different target journals and there are rarely journals that are beneficial for all members of an interdisciplinary research group. Thus, either they do not form in the first place, which suppresses recombination, or they end up agreeing on one set of target journals that is inferior for at least one member of the author team. Again, all things equal, this would lead to lower effort of this team member, and, thus, to a lower overall quality of such a paper. The consequence is straightforward from backward induction: Such interdisciplinary groups do not form in the first place: Either because potential members understand the problem sketched before. Or they have learned from existing interdisciplinary publications that they are methodically weaker, which serves as a signal not to engage with such practices.

Of course, there may exist altruistic reasons why some researchers still engage in such interdisciplinary endeavors. However, these are likely to be outliers. Interdisciplinarity is not incentive compatible for most academics, which is a major hurdle for continued \emph{combinatorical}  growth. Indeed, in a qualitative examination of research practices of young academic economists, \textcite[][p.~329]{Steffy.2023} find that they are ``shying away from interdisciplinary work, avoiding questions that are too specific or could be viewed as only regionally important, and using only those methodologies judged as highly publishable in top journals.'' I consider this as evidence for the theorization above. Furthermore, it is common in economics that authors of a manuscript develop `submission trees,' i.e., elaborate strategies in which journals they should try to get accepted \parencite{Heintzelman.2009}. This emphasizes that the journal rank matters more than finding a suitable journal in the first place

Nevertheless, \textcite{Steffy.2023} distinguish between `discovery' and `delivery' research, where the former is more likely to be combinatorial. They also draw from their investigation that due to the enormous competition of economists for publications in the leading `top 5' journals, there is no single way how to get in there. Put differently, researchers are likely to also follow the `discovery' track in order to create something magnificent that potentially interests these leading journals. \\

\noindent \textbf{Choice Overload:} Lastly, let us turn again towards the fact that academia faces a steady increase in publications and journals. As outlined before, there are substantial search costs involved in entering into a new field. However, this also applies to the `main' fields of a researcher. They also need to stay up-to-date about what is going on in their core fields. Somehow related to increasing search costs, but a distinct mechanism in this context is \emph{choice overload}. It is a behavioral hypothesis that states that an overly large choice set leads to less or less satisfying choices by the choosing individuals. While intuitively compelling, meta-studies have found no clear evidence for the existence of a universal choice overload being present \parencite{Chernev.2015, Scheibehenne.2010}. 

However, they found certain (potential) drivers of choice overload in specific situations: High complexity of the choice set \parencite[][]{Chernev.2015} and the absence of a dominant option and, instead, many non-dominant options being present \parencite{Scheibehenne.2010}. Both are applicable in the case of a growing body of academic journals: Evaluating new journals is not trivial. A researcher has to evaluate whether the outlet is trustworthy or a predatory journal that publishes more or less everything it receives. In light of often similarly sounding titles, this tends to be difficult. Furthermore, even reliable journals need time to establish themselves. Hence, especially for publications in novel outlets, readers must costly assess the quality of the article instead of simply inferring it from the journal's reputation in which it was published. 

If one does not want to incur these evaluation costs or feels overwhelmed by the mass of potentially relevant publications, choice overload may actually lead to a \emph{decrease} in the journals from which a researcher reads and cites papers. Thus, the growth in outlets could actually lead to a smaller choice set of journals a researcher considers worth reading and tracking. This, however, makes recombinations less likely as a researcher gets introduced to fewer ideas. \\

\noindent \textbf{Academic Niches:} There is not only persistent growth in scientific output, this output sorts itself also in an evergrowing set of academic niches as already \textcite{vanRaan.2000} noted when modeling the growth in publications mathematically. However, this `nichefication' bears its own challenges. \textcite{Savin.2024}, for example, examine `degrowth' studies in economics and note that most of them are, in their understanding, rather opinion pieces and do not stand up to the statistical standards established by what is commonly understood as the discipline's `mainstream.' A rebuttal to this critique is often that the mainstream is systematically flawed in its analyses and that it needs different tools. For example, \textcite[][p.~1225]{Koch.2021} state for the subdiscipline of political economy that it \textit{``needs to abandon its anthropocentric ontology and reposition itself in the postgrowth context. This presupposes a break with mainstream economics and an amalgamation with heterodox approaches such as ecological economics, ecofeminism and degrowth.''} 

Put differently, \textcite{Koch.2021} argue in favor of `separating' from the mainstream by using different axioms, paradigms, and, methods. However, this makes different approaches hardly comparable and isolates substrands instead of fostering exchange. The same holds for other heterodox approaches such as Austrian economics in the tradition of Ludwig Mises and Friedrich August Hayek. The latter published one of the most important papers of the discipline in the \textit{American Economic Review}, one of the five top 5 journals and a `cathedral' of the mainstream, namely `\textit{The Use of Knowledge in Society}' \parencite{Hayek.1945}. Nowadays, however, at least two journals exist that are devoted already in their name to Austrian economics, namely the \textit{Review of Austrian Economics} and the \textit{Quarterly Journal of Austrian Economics}. Such an approach, however, diverts `Austrian' concepts and methods from the mainstream to the niche. This can become (which most likely already happened) a self-enforcing process: The more papers diverging from the mainstream and leaning towards a specific strand of the literature are directed to niche journals, the less likely it is that such papers occur in the big general interest journals. But if someone still submits to such a journal, it becomes more likely that the editors do not consider it a good fit for their journal and rather suggest sending it to a more specific (niche) journal. 

This mechanism applies not only to different ideological traditions of economic thinking but also to subfields of the discipline. After a notable increase in interest (namely citations) of the economics discipline in papers dealing with sports, the \textit{Journal of Sports Economics} was established in 2000 \parencite{Kahane.2000}. The editors particularly justify it by the grown interest. However, providing this substrand a distinct `home' also provided papers dealing with sports not only a `safe harbor' in which authors do not need to argue why they examine sports data -- simply because that is the purpose of the journal. It also arguably erected a barrier to entry into general interest journals as there was now a specific journal dedicated to sports economics. In other words, dedicated `niche' journals make it easier for authors to get into these outlets but, at the same time, harder to get into general interest journals that offer visibility not only within but across subdisciplines. This artifact, however, is also likely to weaken the arguments in niche journals, as they do not need to hold up against the criticism that it is not widely applicable. While such a critique sometimes may hinder innovative research to be published, it can actually strengthen a paper as the arguments get sharper and more elaborated if they have to attract a more diverse audience than just a small subdiscipline, in which many readers are often colleagues or even collaborators or coauthors. 

Returning to the combinatorial growth model of \textcite{Koppl.2023}, how can we relate the emergence of academic niches to it? In light of the discussed challenges, the `nichefication' of economics, which is reflected by most academic disciplines, rather leads to a \emph{decombination} of ideas and methods instead of a \emph{recombination} as \textcite{Koppl.2023} observe it for the technosphere. Hence, the growth of niches coincides with less visibility of different approaches as they get diverted into subfield outlets. This narrows the methodical and idea-based corridor of the main general interest journals. At the same time, it may weaken the arguments of the papers appearing in the niche journals as they do not need to be as convincing given there is a high baseline interest in a paper addressing a narrow subfield. This \emph{decombination} makes combinatorial growth in academic research less likely.

\begin{comment}
\FloatBarrier
\section{Niches in Academic Research}
\label{sec.niches}

>> Topic Monopolies instead of interdisciplinarity -- many separate (niche) islands instead of interdisciplinarity

- weniger Überblick über das Feld, Ausdifferenzierung der Subdisziplinen

- Choice overload -- rückkehr zu sehr wenigen Topjournals

- Reputation als Schlüsselfaktor, der für Technologie nicht gilt
\end{comment}

\section{Topics in the Top 5 Economics Journals}
\label{sec.top5}

As mentioned earlier, the top 5 journals are the gatekeepers and clocks of progress in the economics profession. To understand why combinatorial growth may not be possible in academia other than in the technosphere, one has to look at these outlets. To do so, I investigate the content of 8,649 publications in the \textit{American Economic Review, Econometrica}, the \textit{Journal of Political Economy}, the \textit{Quarterly Journal of Economics}, and the \textit{Review of Economic Studies}. To handle this number of publications, I worked with the abstracts of these papers. Of course, doing so implies a loss in information. However, abstracts are purposefully designed to capture the core content of a paper. By that, it is the ideal proxy to conduct a tractable content analysis. 

The data is obtained from the academic publication database Scopus using \emph{pybliometrics} by \textcite{Rose.2019}. The date of the gathering was 6 August 2024. Initially, I could access 11,792 publications. However, for approximately 1/4 of them (3,143), no abstract was available, which excludes them from my analysis. Based on the remaining 8,649 publications, I set up a structural topic model as developed by \textcite{Roberts.2013, Roberts.2019}. It is a method of unsupervised machine learning in which an expectation-maximization (EM) algorithm elicits the underlying themes called topics from a corpus of textual documents. In the present case, the corpus consists of 8,649 abstracts from the top 5 journals in economics, published in the years 1990 - 2023. 

The structural topic model not only allows topics to be correlated with each other but also allows for the incorporation of what is understood in regressions as covariates. In the present case, I include the five journals as well as the publication year as such variables. This leads to 
\begin{align*}
    \theta_d \sim journal_d + year_d 
\end{align*}
Here, the index $d$ captures the document, i.e., the respective paper abstract. $\theta$ is the topic vector that contains the prevalence (in percent) of all $K$ topics for each document. This brings us to the actual choice of the number of topics $K$. Here, it is crucial to acknowledge that this choice does not only have a purely quantitative dimension \parencite{Grimmer.2013}. Rather, one should also carefully evaluate qualitatively the computed topics to which extent they make sense and to which extent the number of topics to be computed is reasonable. I adopt a two-step approach, in which I first choose a set of potential topics based on two quantitative measures, namely \emph{semantic coherence} and \emph{exclusivity}.\footnote{Both measures have diverging mechanisms, which makes it interesting to compare them jointly. While exclusivity increases when more words occur \emph{exclusively} in one topic (instead of several topics), semantic coherence captures how often words that are most likely for a specific topic also co-occur when a high prevalence of this specific topic is measured \parencite{Weston.2023}. The measures are antagonistic as with a higher value of $K$, words usually become more exclusive as topics become more precise -- up to the point where the topics become meaninglessly granular. In that case, exclusivity should decrease again. In contrast, semantic coherence usually decreases in $K$ because, with more granular topics, it is less likely that the words within a topic also co-occur in a specific text \parencite{Schmal.2023}.} In the second step, I qualitatively evaluate the set of pre-selected topics in order to choose one to work with. 

\begin{figure}[htbp]
    \centering
    \includegraphics[width=0.8\linewidth]{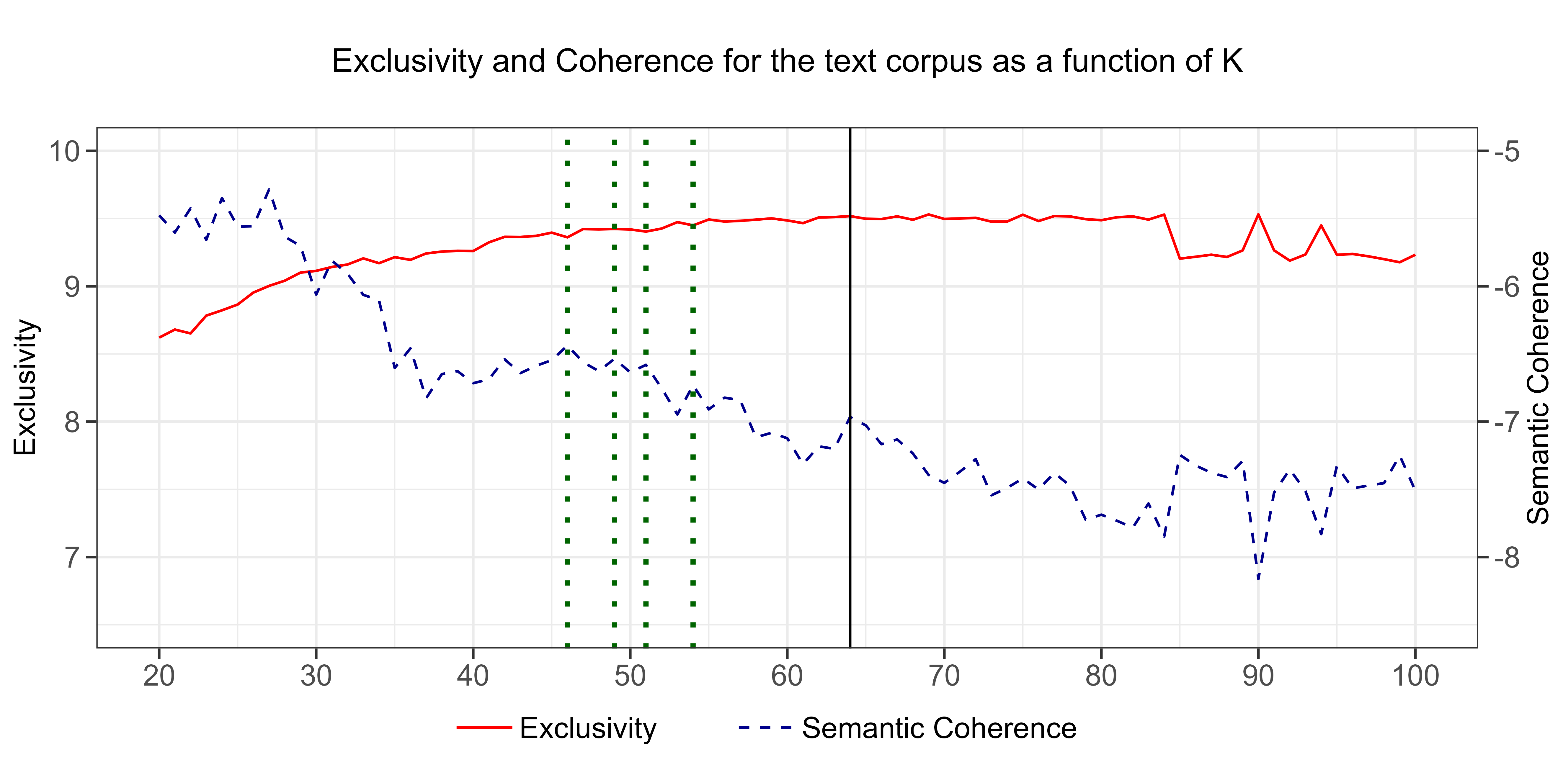}
    \fnote{\footnotesize Values for semantic coherence (SE) and exclusivity (EX) linearly transformed: $\widehat{SE} = \frac{SE}{10}+20$; $\widehat{EX}=(EX-8)\cdot5$, where $\widehat{SE}$ and $\widehat{EX}$ represent the depicted values and $SE$ and $EX$ represent the actual values.}
    \caption{Identification of the optimal number of topics $K^*$}
    \label{fig:optimal-K}
\end{figure}

To do so, I iteratively computed the measures for semantic coherence and exclusivity for $K \: \in [20, 100]$. Figure \ref{fig:optimal-K} depicts this. Looking at both measures jointly is convenient because they have countervailing mechanisms: While an increasing number of topics makes the individual topics less coherent as they become more granular, the topics become more precise, which is, broadly speaking, captured by exclusivity. I use $K=20$ as the lower limit because it is unlikely that some eight thousand articles over the span of more than 30 years contain fewer topics. In light of that, even 20 appears too narrow. In Figure \ref{fig:optimal-K}, I highlight five potentially suitable values for K, namely $\{46,49,51,54,64\}$. This is because at these points, $\Delta SE_{K,K-1}>0$ holds, which is notable in a global environment of decreasing coherence. Of course, there exist more values of K, for which the previously defined condition holds. However, at these five points, we observe notable spikes. 

After carefully evaluating the different sets of topics, I opt to use the highest value of $K=K^*=64$ in my structural topic model. It offers a wide range of topics and is sufficiently granular but still tractable. Tables \ref{tab:topics-1} and \ref{tab:topics-2} present the 64 topics and their keywords.\footnote{I abstain from labeling the topics following a suggestion of \textcite{Mimno.2011} who note that labels may obfuscate the understanding of topics or may bias the interpretation and let readers overlook substantial within-topic variation.} 

\begin{figure}[htbp]
    \centering
    \includegraphics[width=0.8\linewidth]{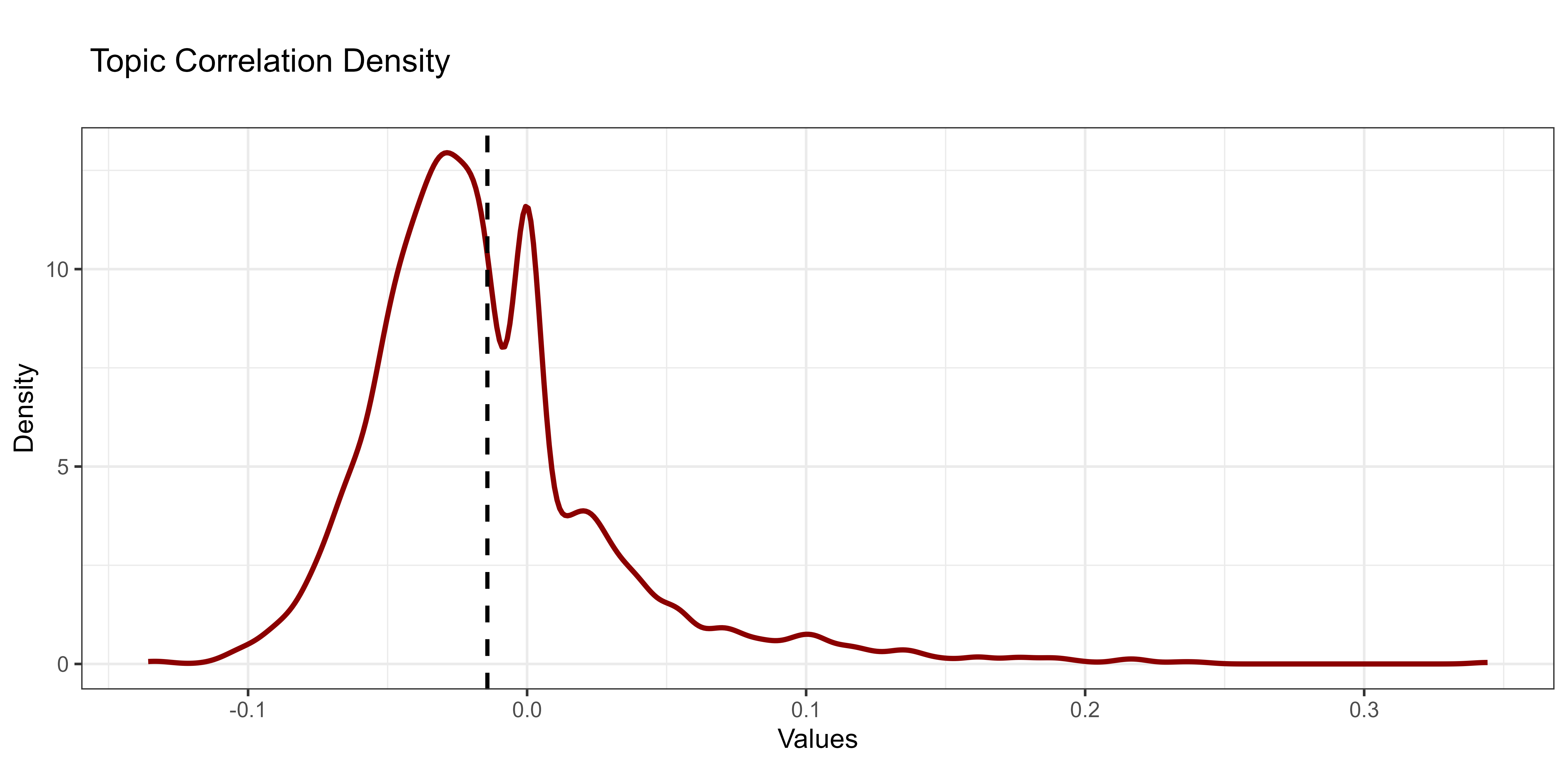}
    \fnote{\footnotesize The dashed line depicts the mean of -0.0142.}
    \caption{Density of correlations across topics}
    \label{fig:corr_dens}
\end{figure}

As structural topic modeling allows for topics to be correlated, it is vital to look at them. Figure \ref{fig:corr_dens} is a density plot for all bilateral correlations, i.e., 64$\times$64 excluding their `own' correlations, which are set to 1 and removed from the dataset. It becomes apparent that correlations approximately follow a normal distribution except for the spike at 0.0. Looking at the descriptive statistics of the distribution of topic correlations shown in Table \ref{tab:desc_stat_corr}, it becomes apparent that they are closely centered around zero, with both mean and median located on the negative side. 

\begin{figure}
    \centering
    \includegraphics[width=0.8\linewidth]{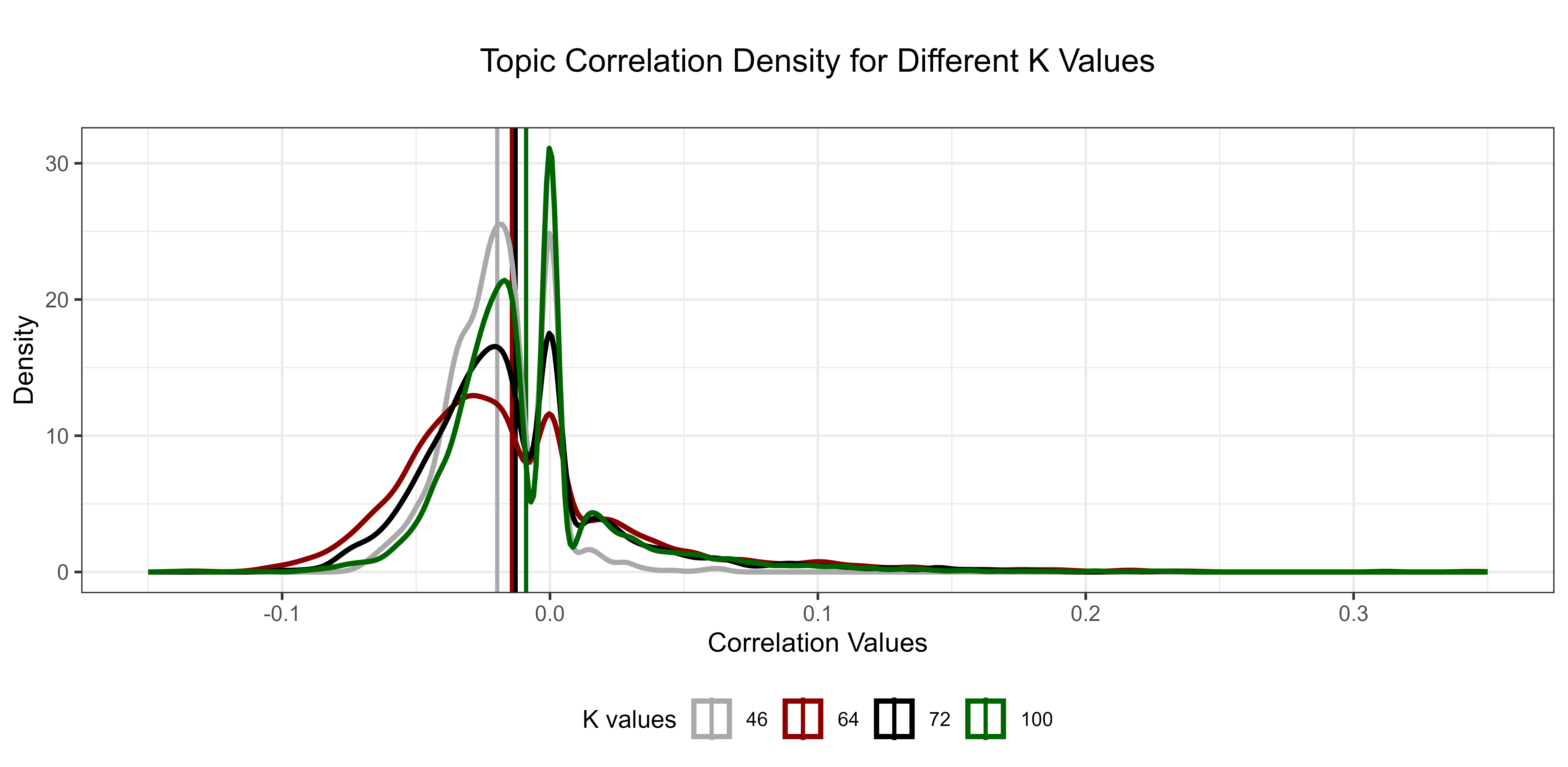}
        \fnote{\footnotesize The solid vertical lines depict the mean values for each of the four distributions. $\mu(K_{46}) = -0.0197,\: \mu(K_{64}) = -0.0142, \:\mu(K_{72}) = -0.0128, \:\mu(K_{100}) = -0.0089.$ $K_{46}$ and $K_{72}$ have been chosen as comparisons based on the optimal K search. $K_{100}$ has been chosen as it captures the highest number of topics considered in the present study. The dark red line of $K_{64}$ reflects is exactly the density plot presented in Fig.~\ref{fig:corr_dens} beforehand.}
    \caption{Density of Topic Correlations for Varying K Values}
    \label{fig:cor_dens_rob}
\end{figure}

This statistical artifact is not the result of a selective choice of the number of topics $K$. Figure \ref{fig:cor_dens_rob} demonstrates that for $K \in \{46, 64^*, 76, 100\}$, correlations are also centered around zero. If, at all, the distribution for $K^*=64$ is more evenly distributed than other values, as one can draw from the additional plot. It underlines the robustness of the finding that topics in the leading top 5 journals are hardly related to each other. 

\begin{table}[htbp]
    \centering
    \begin{tabular}{cccc|c}
    \toprule
    \multicolumn{5}{c}{\textbf{\emph{Topic Correlations}}} \\
    \midrule
        Minimum & Mean & Median & Maximum & Std.~Dev. \\
       -0.1358 & -0.0142 & -0.0220 & 0.3442 & 0.0457 \\
        \bottomrule
    \end{tabular}
    \caption{Descriptive Statistics of Topic Correlations}
    \label{tab:desc_stat_corr}
\end{table}

Looking at the ten highest mutual correlations, which are all positive, we can identify several nets in Table \ref{tab:topic_corrs1}: Topics 29, 31, 40, and 49 are in the subfield of econometrics, where the latter addresses, especially structural models, which are nowadays popular, e.g., in industrial organization. Consequently, topic 52, which is comparatively strongly correlated with the aforementioned topic 49, dynamic approaches and solution algorithms, which also relate to structural models. Together with these two topics, topics 3, 6, 22, and 6, which encompass traditional industrial economics themes, such as the firm, market-entry, and consumer surplus, form the industrial organization cluster.  The third set is formed by topics 12 and 38, which address choice sets and choices under uncertainty. Lastly, topics 36 and 42 capture mechanism and contract design.  

\begin{table}[htbp]
    \centering
    \begin{tabular}{c|c}
    \toprule
    Topic Pair & Correlation \\
    \midrule
      29 - 31   & 0.3442 \\
      29 - 40 & 0.2406 \\
      12 - 38 & 0.2329 \\
      31 - 40 & 0.2200 \\
      6 - 22 &  0.2182 \\
      31 - 49 & 0.2151 \\
      49 - 52 & 0.2117 \\
      29 - 49 & 0.1993 \\
      3 - 61 & 0.1917 \\
      36 - 42 & 0.1912 \\
      \bottomrule
    \end{tabular}
    \caption{Topic Pairs with the Highest Correlation}
    \label{tab:topic_corrs1}
\end{table}
 However, these are only 10 out of 2016 correlation pairs. As we can draw from the characteristics of the correlation distribution, most topics are uncorrelated with each other. What does this tell us about combinatorial growth in the researchsphere of economics, the main question of this paper?

 One can recombine not only methods but also ideas, concepts, themes, or approaches in a narrow or broad sense, such as the topics computed by the present machine learning model covering a wide range of different themes. Thus, correlations across topics are a strong indicator of the extent to which recombination takes place. Considering the vast number of uncorrelated topics, it appears that the five leading journals in academic economics islands instead of networks exist. This is supported by the distribution of topic prevalences depicted in Figure \ref{fig:prob_dens} below. As the structural topic model computes a topic prevalence for all 64 topics among all 8,649 papers, we get a total of 553,536 prevalence values. Figure \ref{fig:prob_dens} shows one massive spike around zero. The rest of the distribution is absolutely flat relative to that. Furthermore, the theoretical mean of 0.015625, which arises from $\frac{1}{K^*}$, exceeds substantially the vast majority of actual topic prevalences. 

\begin{figure}
    \centering
    \includegraphics[width=0.7\linewidth]{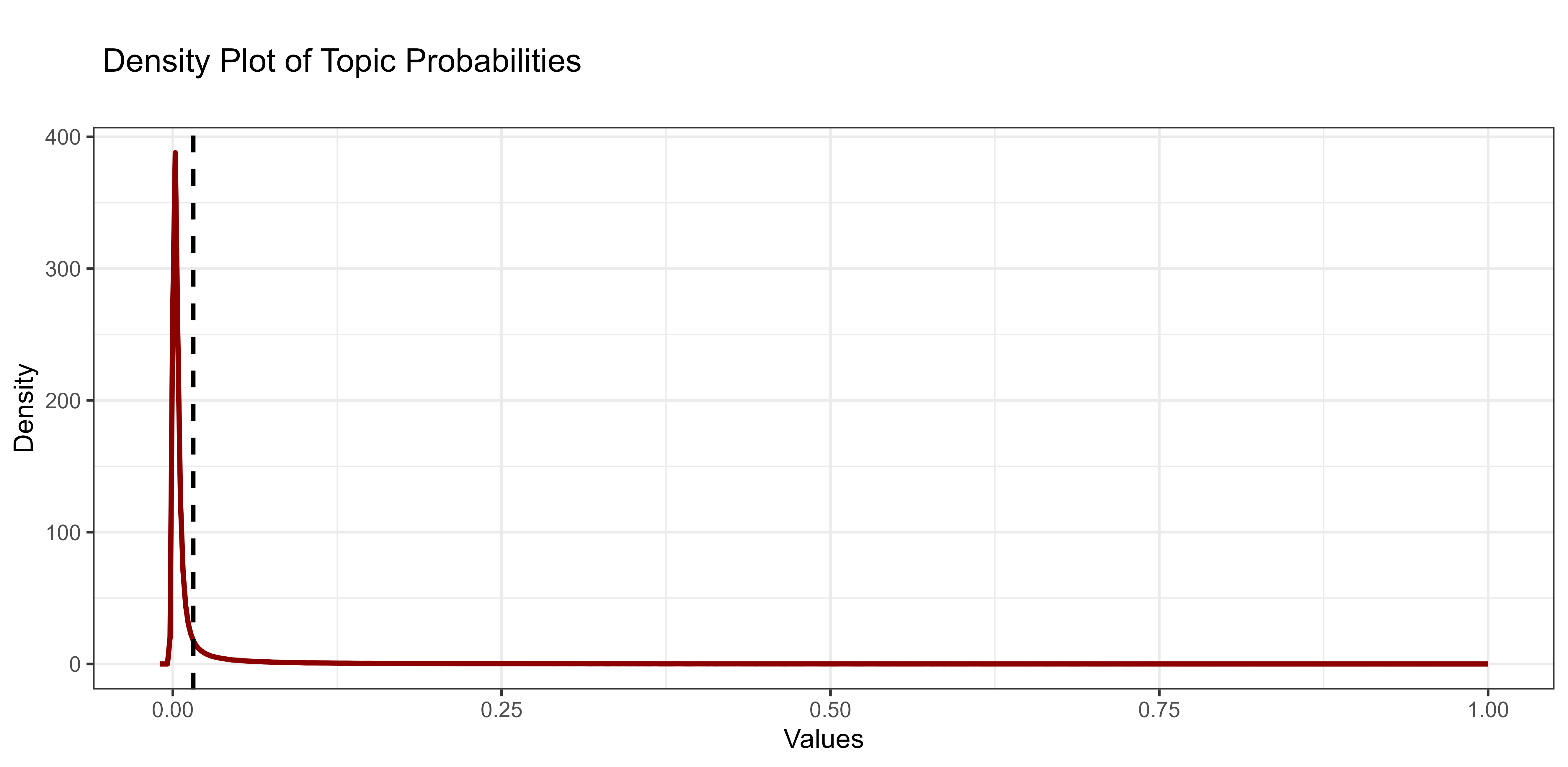}
     \fnote{\footnotesize The dashed line depicts the mean topic probability of 0.0156.}
    \caption{Density Plot for Topic Correlations}
    \label{fig:prob_dens}
\end{figure}

Put differently, in many cases, only very few topics are represented in a paper, while most topics are virtually non-existent. At first sight, it is reasonable that a paper addressing microeconomic questions hardly features macroeconomic themes. But only at first sight. The whole idea of recombinations is the (at first glance) unusual combination of several topics. In order to get combinatorial growth, we would need to see a less severely spiked distribution but rather a bulk of prevalence in the range of 5-20\%. In such a world, the top 5 papers, i.e., the intellectual forefront of the whole discipline, would combine several topics, be it methodical, idea-wise, or something else. In contrast, we observe isolated topics, and each paper is dominated by very few of these isolated topics. This is a strong contra-indicator for combinatorial growth in the academic field of economics.

\section{Discussion and Conclusion}
\label{sec.conc}
%\section{The Researcher as Academic Entrepreneur?}
%\label{sec.kirzner}

\textcite{Wang.2017} analyze that novel research more often than others is cited by foreign disciplines, and occurs more often in strong but not the leading academic journals, i.e., journals with a lower impact factor Furthermore, while often highly cited in the long run, they perform weaker when using shorter citation window metrics. Only in the long run will novel research receive more recognition, e.g., by higher citation rates. In addition, \textcite{Boudreau.2016} find that reviewers evaluate grant proposals less positively if they are more novel. As researchers in economics but also many other disciplines need to publish to not perish, it is highly relevant in which journals they get their work published in \parencite[see, e.g.,][]{Heckman.2020, Schmal.2023RP}. Based on the aforementioned work and the arguments put forward in the previous chapters, it is unlikely that it pays off to work interdisciplinary and recombine existing ideas, concepts, and methods into something really innovative and novel.

It is rather utility maximizing -- especially as long as researchers did not get tenured positions -- to work on `more of the same.' Even if they undoubtedly conduct such research often on an extremely sophisticated level, it is unlikely that such a modus vivendi will trigger combinatorial growth in the way the world saw it in technology \parencite{Koppl.2023}. It is not only a missed chance but is directly relevant to many industrialized countries. Based on the wealth created by the combinatorial growth of the technosphere, a large higher education sector has emerged that is often paid by taxes. A low level of novelty and innovation is a major hurdle to continued growth. However, it can become a competitive advantage once the policymakers in science and higher education alter the career mechanisms in a way that encourages sailing into unchartered waters. 

Lastly, the structural topic model of the contents of the papers in the top 5 journals in economics, the\textit{ American Economic Review, Econometrica,} the \textit{Journal of Political Economy,} the\textit{ Quarterly Journal of Economics,} and the \textit{Review of Economic Studies}, demonstrates a low level of topic correlations that are not caused by a selective choice of the number of topics $K$. Rather, plenty of isolated topic islands seem to exist. Altogether, the evidence presented in this paper cannot detect a high chance for combinatorial growth that reflects the tremendous growth path of the technosphere \parencite{Koppl.2023}. \\
For the `researchsphere,' I rather observe \emph{decombinations} instead of \emph{recombinations}. 

%\vspace{5mm}
%\pagebreak
\begin{singlespace}
\printbibliography
\addcontentsline{toc}{section}{\protect\numberline{}References}
\end{singlespace}
%\pagebreak

\vspace{10mm}
\newpage
\section*{Appendix}
\label{sec.app}
\renewcommand{\thefigure}{F\arabic{figure}}
\setcounter{figure}{0}
\renewcommand{\thetable}{A\arabic{table}}
\setcounter{table}{0}
\FloatBarrier
\vspace{-5mm}
\begin{singlespace}
\subsection*{Corpus Preprocessing}

Preparing the textual content for the unsupervised machine-learning technique of topic modeling is a crucial task that may substantially affect results \parencite{Denny.2018}. In the present case, I performed a two-step approach. First, I manually executed a punctual cleaning of the abstracts. In the second step, I conduct automatized preprocessing based on the packages provided in \verb|R|, namely \verb|quanteda| \parencite{quanteda.2018} and \verb|STM| \parencite{stm.2019}. 

First, I remove copyright labels at the end of abstracts usually listed by Scopus. This encompasses the following expressions, including spelling and grammar errors, for example, `All right reserved' where an `s' is missing at the end of `right.'
\begin{itemize}
    \item All rights reserved.
    \item The University of Chicago. All rights reserved.
    \item © The Author(s)
    \item Published by Oxford University Press on behalf of The Review of Economic Studies Limited
    \item by the President and Fellows of Harvard College and the Massachusetts Institute of Technology.
    \item Published by Oxford University Press on behalf of
    \item AEA. The American Economic Association is hosted by Vanderbilt University.
\end{itemize}
Furthermore, I manually stem the terms `equilibrium' and `equilibria' to `equilibri.' Usually, the task of word stemming is performed independently by the named packages. However, these tools rely on libraries in the backend, which, apparently, do not list the term equilibrium and its flections. Not stemming would result in the problem that the topic model algorithm would consider equilibrium and equilibria to be separate, independent words instead of singular and plural of the same term. 

In the second step, I clean the data technically in \verb|R|. To do that, I remove stopwords based on default stopword lists and add minor errors as well as words related to academic publishing that do not add any insights, for example, `paper,' or `publish', which often relates to other publications without saying anything about the content. Aside from removing stopwords, I stem words as mentioned. I do this in order to avoid double counting and distorted word clustering. Lastly, I also remove punctuation.

\FloatBarrier
\begin{table}
\scriptsize
\begin{tabular}{l|ll}
\toprule
\hline
% Top Words:                                                               \\
Topic 1 & \textbf{Highest Prob:} polici, social, network, interact, monetari, public,   central     \\
& \textbf{FREX:} polici, network, social, interact, target, monetari, norm                  \\
% &  Top Words:                                                               \\
\hline
Topic 2& \textbf{Highest Prob:} present, shown, error, formula, correct, discuss,   equat          \\
& \textbf{FREX:} shown, note, correct, formula, discuss, error, present                     \\
% &  Top Words:                                                               \\
\hline
Topic 3 & \textbf{Highest Prob:} consum, surplus, good, custom, attent, qualiti, pay                \\
& \textbf{FREX: }consum, surplus, attent, custom, bundl, tast, inattent                     \\
% &  Top Words:                                                               \\
\hline
Topic 4 & \textbf{Highest Prob:} firm, manag, perform, profit, compens, pay, corpor                 \\
& \textbf{FREX:} manag, manageri, ceo, firm, sharehold, compens, corpor                     \\
% & Top Words:                                                               \\
\hline
Topic 5 & \textbf{Highest Prob: }labor, work, suppli, hour, time, forc, elast                       \\
& \textbf{FREX: }hour, work, labor, disabl, suppli, week, leisur                            \\
% & Top Words:                                                               \\
\hline
Topic 6 & \textbf{Highest Prob:} product, industri, output, input, produc, manufactur,   sector     \\
&\textbf{ FREX:} product, industri, manufactur, input, realloc, plant,   output             \\
% &  Top Words:                                                               \\
\hline
Topic 7 & \textbf{Highest Prob:} state, unit, steadi, coalit, capac, stabl, world                   \\
& \textbf{FREX: }state, steadi, coalit, unit, capac, approv, kingdom                        \\
% & Top Words:                                                               \\
\hline
Topic 8 & \textbf{Highest Prob: }rate, interest, exchang, real, currenc, low,   monetari            \\
& \textbf{FREX: }currenc, interest, rate, forward, real, pariti, exchang                    \\
% & Top Words:                                                               \\
\hline
Topic 9 & \textbf{Highest Prob:} measur, use, indic, index, survey, method, robust                  \\
& \textbf{FREX:} measur, index, methodolog, indic, elicit, survey, proxi                    \\
% & Top Words:                                                              \\
\hline
Topic 10 & \textbf{Highest Prob:} predict, bias, belief, forecast, discrimin, test,   data           \\
& \textbf{FREX:} forecast, predict, discrimin, bias, belief, minor, analyst                 \\
% &  Top Words:                                                              \\
\hline
Topic 11 & \textbf{Highest Prob:} invest, growth, capit, human, accumul, develop, per                \\
& \textbf{FREX:} invest, capit, human, growth, accumul, capita, gdp                         \\
% & Top Words:                                                              \\
\hline
Topic 12 & \textbf{Highest Prob:} prefer, choic, set, function, individu, altern,   order            \\
& \textbf{FREX:} choic, prefer, monoton, function, axiom, order, domin                      \\
% & Top Words:                                                              \\
\hline
Topic 13 & \textbf{Highest Prob:} reform, govern, econom, regim, chang, system,   centuri            \\
& \textbf{FREX:} regim, europ, reform, centuri, deficit, soviet, china                      \\
% & Top Words:                                                              \\
\hline
Topic 14 & \textbf{Highest Prob:} auction, bid, valu, bidder, revenu, privat, object                 \\
& \textbf{FREX:} auction, bidder, bid, first-pric, second-pric, vickrey,   curs             \\
% &  Top Words:                                                              \\
\hline
Topic 15 & \textbf{Highest Prob:} regul, cost, pollut, estim, emiss, environment, air                \\
& \textbf{FREX:} pollut, air, emiss, electr, regul, energi, fuel                            \\
% & Top Words:                                                              \\
\hline
Topic 16 & \textbf{Highest Prob:} game, equilibri, player, payoff, strategi, bargain,   nash         \\
& \textbf{FREX:} player, nash, game, payoff, bargain, strategi, repeat                      \\
% & Top Words:                                                              \\
\hline
Topic 17 & \textbf{Highest Prob:} polit, vote, voter, elect, parti, candid, govern                   \\
& \textbf{FREX: }voter, elect, politician, elector, vote, candid, democrat                  \\
% & Top Words:                                                              \\
\hline
Topic 18 &\textbf{ Highest Prob: }save, contribut, subsidi, plan, age, retir,   individu             \\
& \textbf{FREX: }save, retir, subsidi, contribut, subsid, plan,   precautionari             \\
% & Top Words:                                                              \\
\hline
Topic 19 & \textbf{Highest Prob:} agricultur, right, land, farmer, data, properti,   farm            \\
& \textbf{FREX: }farmer, farm, land, agricultur, crop, misalloc, right                      \\
% & Top Words:                                                              \\
\hline
Topic 20 & \textbf{Highest Prob:} institut, war, ethnic, conflict, group, histor,   cultur           \\
& \textbf{FREX:} ethnic, cultur, war, civil, religi, trust, histor                          \\
% & Top Words:                                                              \\
\hline
Topic 21 & \textbf{Highest Prob:} price, demand, cost, seller, buyer, profit, suppli                 \\
& \textbf{FREX: }seller, inventori, buyer, demand, collus, monopoli,   oligopoli            \\
% & Top Words:                                                              \\
\hline
Topic 22 & \textbf{Highest Prob: }firm, export, foreign, markup, exit, data, size                    \\
& \textbf{FREX: }export, multin, markup, firm-level, exit, foreign, fdi                     \\
% & Top Words:                                                              \\
\hline
Topic 23 & \textbf{Highest Prob: }school, student, educ, colleg, score, teacher, high                \\
& \textbf{FREX: }school, teacher, student, colleg, score, attend, grade                     \\
% & Top Words:                                                              \\
\hline
Topic 24 & \textbf{Highest Prob: }worker, employ, wage, job, unemploy, search, labour                \\
& \textbf{FREX: }unemploy, job, employ, search, worker, hire, vacanc                        \\
% & Top Words:                                                              \\
\hline
Topic 25 & \textbf{Highest Prob:} technolog, innov, adopt, research, new, patent,   spillov          \\
& \textbf{FREX:} patent, innov, inventor, technolog, adopt, spillov, invent                 \\
% & Top Words:                                                              \\
\hline
Topic 26 & \textbf{Highest Prob: }inform, privat, learn, signal, observ, asymmetr,   public          \\
& \textbf{FREX:} inform, signal, learn, uninform, asymmetr, privat,   disclosur             \\
% & Top Words:                                                              \\
\hline
Topic 27 & \textbf{Highest Prob:} trade, countri, intern, world, global, develop,   good             \\
& \textbf{FREX:} intern, countri, trade, tariff, global, liber, bilater                     \\
% & Top Words:                                                              \\
\hline
Topic 28 & \textbf{Highest Prob:} payment, servic, care, hospit, patient, health,   qualiti          \\
& \textbf{FREX:} physician, payment, hospit, servic, care, patient, medicar                 \\
% & Top Words:                                                              \\
\hline
Topic 29 & \textbf{Highest Prob: }condit, function, paramet, econometr, distribut, bound,   set      \\
& \textbf{FREX: }interv, confid, bound, econometr, quantil, moment, paramet                 \\
% & Top Words:                                                              \\
\hline
Topic 30 & \textbf{Highest Prob:} women, men, gender, femal, marriag, fertil,   children             \\
& \textbf{FREX: }marriag, femal, women, marri, gender, men, sexual                          \\
% & Top Words:                                                              \\
\hline
Topic 31 & \textbf{Highest Prob:} estim, method, sampl, consist, regress, data,   variabl            \\
& \textbf{FREX:} nonparametr, semiparametr, squar, estim, covari, regressor,   regress      \\
% & Top Words:     \\
\hline
Topic 32 & \textbf{Highest Prob:} crime, effect, crimin, find, ownership, law,   increas             \\
& \textbf{FREX:} crimin, crime, polic, violent, incarcer, judg, arrest          \\
%\bottomrule
\end{tabular}
\caption{List of Topics, Part 1}
\label{tab:topics-1}
\end{table}

\begin{table}[htbp]
    \centering
    \scriptsize
    \begin{tabular}{l|ll}
\toprule
% Top Words:                                                              \\
Topic 33  & \textbf{Highest Prob:} decis, make, action, maker, choos, individu,   conflict            \\
& \textbf{FREX:} decis, maker, action, deleg, quo, herd, make                               \\
% Top Words:                                                              \\
\hline
Topic 34 & \textbf{Highest Prob: }news, media, onlin, review, corrupt, result, data                  \\
& \textbf{FREX: }onlin, corrupt, media, news, proposit, review, outlet                      \\
%Top Words:                                                              \\
\hline
Topic 35 & \textbf{Highest Prob:} price, chang, find, good, retail, advertis, purchas                \\
& \textbf{FREX:} retail, durabl, advertis, store, brand, price, purchas                     \\
%Top Words:                                                              \\
\hline
Topic 36 & \textbf{Highest Prob:} agent, optim, mechan, alloc, implement, type,   design             \\
& \textbf{FREX:} agent, mechan, implement, princip, alloc, compat, contest                  \\
%Top Words:                                                              \\
\hline
Topic 37 & \textbf{Highest Prob:} shock, aggreg, respons, cycl, inflat, busi, output                 \\
& \textbf{FREX: }shock, fluctuat, cycl, aggreg, recess, busi, macroeconom                   \\
%Top Words:                                                              \\
\hline
Topic 38 & \textbf{Highest Prob:} risk, util, expect, uncertainti, prefer, avers,   subject          \\
& \textbf{FREX: }avers, ambigu, util, uncertainti, gambl, lotteri, represent                \\
%Top Words:                                                              \\
\hline
Topic 39 & \textbf{Highest Prob: }local, citi, region, locat, area, popul, migrat                    \\
& \textbf{FREX: }citi, migrat, locat, spatial, region, agglomer, local                      \\
%Top Words:                                                              \\
\hline
Topic 40 & \textbf{Highest Prob:} test, statist, asymptot, hypothesi, power, bootstrap,   base       \\
& \textbf{FREX:} null, bootstrap, statist, test, root, cointegr, hypothesi                  \\
%Top Words:                                                              \\
\hline
Topic 41 & \textbf{Highest Prob:} financi, debt, constraint, default, financ, borrow,   govern       \\
& \textbf{FREX: }debt, default, sovereign, financ, financi, matur, collater                 \\
%Top Words:                                                              \\
\hline
Topic 42 & \textbf{Highest Prob: }contract, incent, effort, commit, cost, project,   moral           \\
& \textbf{FREX: }contract, moral, incent, project, renegoti, effort, hazard                 \\
%Top Words:                                                              \\
\hline
Topic 43 & \textbf{Highest Prob:} subject, experi, communic, coordin, cooper, experiment,   behavior \\
& \textbf{FREX: }communic, coordin, cooper, laboratori, punish, experiment,   subject       \\
%Top Words:                                                              \\
\hline
Topic 44 & \textbf{Highest Prob:} rule, reserv, complex, optim, constitut, must,   propos            \\
& \textbf{FREX:} rule, complex, constitut, reserv, prioriti, unanim, block                  \\
%Top Words:                                                              \\
\hline
Topic 45 & \textbf{Highest Prob:} earn, gap, mobil, data, wage, educ, black                          \\
& \textbf{FREX:} earn, mobil, gap, black, immigr, young, longitudin                         \\
%Top Words:                                                              \\
\hline
Topic 46 & \textbf{Highest Prob:} tax, incom, taxat, rate, optim, govern, increas                    \\
& \textbf{FREX:} tax, taxat, evas, ramsey, taxpay, multipli, wedg                           \\
%Top Words:                                                              \\
\hline
Topic 47 & \textbf{Highest Prob:} process, time, stochast, converg, continu, compon,   stationari    \\
& \textbf{FREX:} stochast, process, markov, stationari, converg, continuous-tim,   compon   \\
%Top Words:                                                              \\
\hline
Topic 48 & \textbf{Highest Prob:} hous, children, effect, exposur, famili, outcom,   home            \\
& \textbf{FREX:} hous, neighborhood, childhood, segreg, exposur, mortgag,   home            \\
%Top Words:                                                              \\
\hline
Topic 49 & \textbf{Highest Prob:} model, structur, heterogen, distribut, develop, observ,   allow    \\
& \textbf{FREX:} heterogen, structur, model, unobserv, framework, endogen,   identifi       \\
%Top Words:                                                              \\
\hline
Topic 50 & \textbf{Highest Prob:} bank, credit, liquid, loan, lend, crisi, boom                      \\
& \textbf{FREX:} bank, deposit, credit, loan, lend, boom, crisi                             \\
%Top Words:                                                              \\
\hline
Topic 51 & \textbf{Highest Prob:} equilibri, effici, exist, match, economi, market,   general        \\
& \textbf{FREX:} equilibri, multipl, exist, match, pareto, effici, ineffici                 \\
%Top Words:                                                              \\
\hline
Topic 52 & \textbf{Highest Prob:} dynam, approach, problem, comput, use, solut,   algorithm          \\
& \textbf{FREX:} approach, algorithm, comput, dynam, static, solut, solv                    \\
%Top Words:                                                              \\
\hline
Topic 53 & \textbf{Highest Prob:} effect, experi, random, treatment, group, control,   evalu         \\
& \textbf{FREX:} treatment, random, evalu, experi, effect, treat, field                     \\
%Top Words:                                                              \\
\hline
Topic 54 & \textbf{Highest Prob:} transfer, parent, fund, grant, children, support,   famili         \\
& \textbf{FREX:} grant, transfer, altruism, altruist, fund, prosoci, parent                 \\
%Top Words:                                                              \\
\hline
Topic 55 & \textbf{Highest Prob:} skill, wage, chang, worker, increas, relat, labor                  \\
& \textbf{FREX:} skill, sort, occup, autom, technic, task, skill-bias                       \\
%Top Words:                                                              \\
\hline
Topic 56 & \textbf{Highest Prob:} asset, price, market, stock, investor, trade,   equiti             \\
& \textbf{FREX:} investor, asset, bubbl, portfolio, equiti, stock, specul                   \\
%Top Words:                                                              \\
\hline
Topic 57 & \textbf{Highest Prob:} theori, behavior, econom, evid, empir, theoret,   term             \\
& \textbf{FREX:} theori, question, peopl, theoret, view, fair, term                         \\
%Top Words:                                                              \\
\hline
Topic 58 & \textbf{Highest Prob:} incom, household, consumpt, wealth, inequ, distribut,   share      \\
& \textbf{FREX:} household, wealth, consumpt, incom, perman, inequ,   expenditur            \\
%Top Words:                                                              \\
\hline
Topic 59 & \textbf{Highest Prob:} margin, particip, extens, intens, cost, chang,   adapt             \\
& \textbf{FREX:} margin, lie, extens, particip, damag, intens, adapt                        \\
%Top Words:                                                              \\
\hline
Topic 60 & \textbf{Highest Prob:} exchang, organ, donor, give, transplant, increas,   gift           \\
& \textbf{FREX:} kidney, transplant, donor, gift, donat, wait, chariti                      \\
%Top Words:                                                              \\
\hline
Topic 61 & \textbf{Highest Prob:} market, welfar, competit, effect, gain, loss, entri                \\
& \textbf{FREX:} welfar, merger, competit, dealer, market, vertic, airlin                   \\
%Top Words:                                                              \\
\hline
Topic 62 & \textbf{Highest Prob: }percent, insur, health, estim, increas, cost, reduc                \\
& \textbf{FREX:} insur, percent, mortal, medicaid, health, drug, annuiti                    \\
%Top Words:                                                              \\
\hline
Topic 63 & \textbf{Highest Prob:} model, return, data, explain, predict, correl,   factor            \\
& \textbf{FREX:} puzzl, return, constant, habit, intertempor, excess,   time-vari           \\
%Top Words:                                                              \\
\hline
Topic 64 & \textbf{Highest Prob:} analyz, differ, investig, use, new, studi, result                  \\
& \textbf{FREX:} analyz, investig, differ, new, unlik, use, sever                     \\
\bottomrule
   \end{tabular}
   \caption{List of Topics, Part 2}
\label{tab:topics-2}
\end{table}
%\end{longtable}

% \newpage

\begin{comment}
    \begin{figure}[htbp]
    \centering
    \includegraphics[width=0.75\linewidth]{corr_aer.png}\\
    \includegraphics[width=0.75\linewidth]{corr_qje.png}\\
    \includegraphics[width=0.75\linewidth]{corr_jpe.png}\\
    \includegraphics[width=0.75\linewidth]{corr_RESTUD.png}
    \caption{Topic Correlations by Journal}
    \label{fig:topic_corr_all}
\end{figure}
\end{comment}

\end{singlespace}

\end{doublespace}
\end{document}